\documentstyle[12pt]{article}
\begin{document}
{\pagestyle{empty}
\renewcommand{\thefootnote}{\fnsymbol{footnote}}
\rightline{May 1997}
\rightline{~~~~~~~~~}
\vskip 1cm
\centerline{\Large \bf Scalar-Tensor Theory of Gravity on $M_4\times \mbox{\boldmath{$Z$}}_2$ Geometry}
\vskip 1cm

\centerline{
  {Akira Kokado\footnote{E-mail address: kokado@kgupyr.kwansei.ac.jp}}}

\centerline{
  {\it Kobe International University, Kobe 655, Japan} }

\vskip .6cm

\centerline{%
  {Gaku Konisi\footnote{E-mail address: konisi@kgupyr.kwansei.ac.jp}},
  {Takesi Saito\footnote{E-mail address: tsaito@jpnyitp.yukawa.kyoto-u.ac.jp}},
  and
  {Yutaka Tada\footnote{E-mail address: tada@kgupyr.kwansei.ac.jp}}}

\centerline{
  {\it Department of Physics, Kwansei Gakuin University,
  Nishinomiya 662, Japan}}

\vskip 1.3cm

\centerline{
  {(Dedicated to the late Prof. Robert H. Dicke)}}

\vskip 0.5cm

\centerline{\bf Abstract}
\vskip 0.2in

 In the Brans-Dicke(BD) theory on $M_{4}\times \mbox{\boldmath{$Z$}}_2$ 
geometry the geometrical meaning of the torsion is clarified. The BD 
theory on $M_{4}\times \mbox{\boldmath{$Z$}}_2$ is rederived by taking
 into account of a new isometry condition.

\vskip 0.4cm\noindent
PACS number(s):04.50.+h, 46.10
\hfil
\vfill
\newpage}
%%%%%%%%%%%%%%%%%%%%%%%%%%%%%% Section 1 %%%%%%%%%%%%%%%%%%%%%%%%%%%%%%%
\newcommand{\iDelta}{{\mit\Delta}}
\newcommand{\iGamma}{{\mit\Gamma}}
\renewcommand{\thesection}{\Roman{section}.}
\renewcommand{\theequation}{\arabic{section}.\arabic{equation}}
\setcounter{equation}{0}
\section{Introduction}

\indent

Recently we have shown\cite{Kokado} that the pure Einstein action on $M_{4}\times \mbox{\boldmath{$Z$}}_2$
geometry exactly leads to the Brans-Dicke (BD) theory\cite{Brans} in 4-dimensional
space-time $M_{4}$, where a scalar field is coupled to gravity. Here the $%
\mbox{\boldmath{$Z$}}_2$ is a discrete space with two points. We have used the geometrical theory on $%
M_{4}\times \mbox{\boldmath{$Z$}}_2$, which was previously proposed by two of authors\cite{Konisi} without
recourse to the noncommutative geometry (NCG) of Connes\cite{Connes}. On this manifold $%
M_{4}\times \mbox{\boldmath{$Z$}}_2$ there are three kinds of Riemann curvature tensors. We
have clarified the geometrical meaning of them.

 In order to calculate the Riemann curvature we should express the affine connections $\iGamma _{LMN}$
in terms of the metric $G_{MN}$ in this space. This space formally can be
regarded as the 5-dimensional Kaluza-Klain like space where the fifth
continuous dimension is replaced by two points. On $M_{4}$ usually one uses
the isometry condition that any inner product of vectors is invariant under
the parallel-transportation of vectors. This leads to the covariant constancy
of the metric $g_{\mu \nu }$%
\begin{equation}
{\partial }_{\lambda }g_{\mu \nu }=\iGamma _{\mu \lambda \nu }+\iGamma _{\nu
\lambda \mu }.
\end{equation}

\noindent
As well known, if the affine connection is symmetric, $\iGamma
_{\mu \lambda \nu }=\iGamma _{\mu \nu \lambda }$, then Eq.(1.1) is used to
express $\iGamma $ in terms of $g_{\mu \nu }$. Even on $M_{4}\times \mbox{\boldmath{$Z$}}_2$ we
can do the same thing. In this case we have two kinds of equations

\begin{equation}
{\partial }_{\lambda }G_{MN}=\iGamma _{M\lambda N}+\iGamma _{N\lambda M},
\end{equation}

\begin{equation}
{\partial }_{r}G_{MN}=\iGamma _{MrN}+\iGamma _{NrM}+
\iGamma _{KrM}\iGamma _{rN}^{K}\triangle ^{r}z,
\end{equation}

\noindent
where $M=(\mu ,r)$, $\partial _{r}$ is a derivative on $\mbox{\boldmath{$Z$}}_2$, and $%
\triangle ^{r}z=z(g+r)-z(g)$ defined below. The point is that Eq.(1.3) is
valid for any $\triangle ^{r}z$ as explained in the text. Since $\partial
_{r}\triangle ^{r}z=-2$, there are contributions from the last term. In the
previous work\cite{Kokado} we have not taken into account of such term. This term is so
important as to eliminate the Riemann tensor of the third type, and to
yield the BD kinetic term.

The purpose of this paper is to reconsider the scalar-tensor theory of
gravity on $M_{4}\times \mbox{\boldmath{$Z$}}_2$, by using the new isometry 
condition (1.3). We also consider the geometrical meaning of torsion on 
$M_{4}\times \mbox{\boldmath{$Z$}}_2$, which has not so for been discussed enough.
To begin we need the equivalence assumption proposed in the previous work.
This is stated as follows: the manifold $M_{4}\times \mbox{\boldmath{$Z$}}_2$ may be regarded
as a pair of $M_{4}$, each at the point $e$ or $r$ on $\mbox{\boldmath{$Z$}}_2$. The physics
on these two $M_{4}$-pieces should be equivalent to each other. We assume
that the equivalence is attained by a limiting process with some parameter $%
\varepsilon $ which tends to zero. As a technical tool to express the
limiting process we introduce a coordinate $z(g)$ on $\mbox{\boldmath{$Z$}}_2$ $(g=e$ or $r)$
such that a difference

\begin{equation}
\triangle ^{r}z(g)=z(g+r)-z(g)\sim \varepsilon 
\end{equation}

\noindent
is proportional to the limiting process parameter $\varepsilon $. Let $f
(z(g))$ be any function on $\mbox{\boldmath{$Z$}}_2$. Since $f(z(g))$ is a linear function
of $z(g)$, one may put

\begin{equation}
f\left( z(g)\right) =A+Bz(g),
\end{equation}

\noindent
where $A$ and $B$ are some constants. From this we find that its Taylor
expansion is cut off only up to the first order $\triangle ^{r}z(g)$

\begin{equation}
f\left( z(g+r)\right) =f\left( z(g)\right) +B\triangle ^{r}z(g).
\end{equation}

\noindent
Now, the equivalence indicates that

\begin{equation}
f\left( x,z(g+r)\right) \longrightarrow f\left( x,z(g)\right) \mbox{
as }\varepsilon \longrightarrow 0,
\end{equation}

\noindent
where the coordinate $x$ on $M_{4}$ is inserted.

In \S 2 we consider the geometrical meaning of torsion. In \S 3 the
isometry condition is required in order to derive Eqs.(1.2) and (1.3). In
\S 4 we calculate three kind of Riemann curvature tensors and derive the BD
theory. The final section is devoted to concluding remarks.
\newpage
%%%%%%%%%%%%%%%%%%%%%%%%%%%%%% Section 2 %%%%%%%%%%%%%%%%%%%%%%%%%%%%%%%
\setcounter{equation}{0}
\section{Torsion on $M_4 \times Z_2$}

\indent

We consider the tangent space $T(p)$ at a point $P$ on $M_{4}\times \mbox{\boldmath{$Z$}}_2$
with local coordinates ($x^{\mu }$,$g$), $x^{\mu }\in M_{4}$ and $g=\{e$%
(unit element), $r\}\in \mbox{\boldmath{$Z$}}_2$. Let the origin of $T(x,g)$ be $O(x,g)$. At a
point $(x^{\mu }+\triangle x^{\mu },g)$ very close to $P$ we have also
another tangent space $T(x+\triangle x,g)$, whose origin is $O(x+\triangle
x,g)$. We consider a mapping of the origin $O(x+\triangle x,g)$ from $%
T(x+\triangle x,g)$ onto $T(x,g)$. The mapped point is denoted by a notation 
$U(x,x+\triangle x,g)O(x+\triangle x,g)$. A covariant difference between $%
U(x,x+\triangle x,g)O(x+\triangle x,g)$ and $O(x,g)$ defines a vector $%
\mbox{\boldmath{$e$}}_{\mu }(x,g)$ on $T(x,g)$

\begin{eqnarray}
\triangle _{x}O(x,g) &=&U(x,x+\triangle x,g)O(x+\triangle x,g)-O(x,g) \nonumber \\
&=&\mbox{\boldmath{$e$}}_{\mu }(x,g)\triangle x^{\mu }.
\end{eqnarray}

In the same way the mapping of the origin $O(x,g+r)$ from $%
T(x,g+r) $ onto $T(x,g)$ is given by the notation $U(x,g,g+r)O(x,g+r)$. 
The covariant difference between the mapped point
and $O(x,g)$ defines another vector $\mbox{\boldmath{$e$}}_{r}(x,g)$ on $T(x,g)$%
\begin{eqnarray}
\triangle _{r}O(x,g) &=&U(x,g,g+r)O(x,g+r)-O(x,g) \nonumber \\
&=&\mbox{\boldmath{$e$}}_{r}(x,g)\triangle ^{r}z(g),
\end{eqnarray}

\noindent
where $z(g)$ is the coordinate corresponding to $g$ and 
\begin{equation}
\triangle ^{r}z(g)=z(g+r)-z(g)
\end{equation}

\noindent
is proportional to the limiting process parameter $\varepsilon $. A set of
vectors 
\begin{equation}
\mbox{\boldmath{$e$}}_{N}(x,g)=\left\{ \mbox{\boldmath{$e$}}_{\mu }(x,g),\mbox{\boldmath{$e$}}_{r}(x,g),g=(e,r)\in \mbox{\boldmath{$Z$}}_2\right\}
\end{equation}

\noindent
supplies a basis on $T(x,g)$. In this paper we do not consider the direct
mapping of the origin $O(x+\triangle x,g+r)$ from $T(x+\triangle x,g+r)$ to $T(x,g)$.

Let us then consider the mapping of the basis $\mbox{\boldmath{$e$}}_{N}(x+\triangle x,g)$ from $%
T(x+\triangle x,g)$ onto $T(x,g)$. The mapped basis which is denoted by $\mbox{\boldmath{$e$}}_{N}^{\
H}(x+\triangle x,g)$ is given by a rotation $H_{\ N}^{M}(x,x+\triangle x,g)$
of $\mbox{\boldmath{$e$}}_{N}(x,g)$%
\begin{equation}
\mbox{\boldmath{$e$}}_{N}^{\ H}(x+\triangle x,g)=\mbox{\boldmath{$e$}}_{M}(x,g)H_{\ N}^{M}(x,x+\triangle x,g).
\end{equation}

\noindent
We define the affine connection $\widehat{\iGamma }_{\ N\mu }^{M}(x,g)$
by 
\begin{equation}
H_{\ N}^{M}(x,x+\triangle x,g)=\delta _{\ N}^{M}+\widehat{\iGamma }%
_{\ N\mu }^{M}(x,g)\triangle x^{\mu }+O(\triangle x)^{2}.
\end{equation}

\noindent
Substituting this into the above equation we have
\begin{equation}
\mbox{\boldmath{$e$}}_{N}^{\ H}(x+\triangle x,g)=\mbox{\boldmath{$e$}}_{N}(x,g)+\widehat{\mbox{\boldmath{$\iGamma$}}}_{N\mu
}(x,g)\triangle x^{\mu }+O(\triangle x)^{2},
\end{equation}

\noindent
where
\begin{equation}
\widehat{\mbox{\boldmath{$\iGamma $}}}_{N\mu }(x,g)\equiv \mbox{\boldmath{$e$}}_{M}(x,g)\widehat{\iGamma }_{\ N\mu
}^{M}(x,g).
\end{equation}

\noindent
This equation defines the covariant difference of $\mbox{\boldmath{$e$}}_{N}(x,g)$ along $M_{4}$%
\begin{equation}
\triangle _{x}\mbox{\boldmath{$e$}}_{N}(x,g)\equiv \mbox{\boldmath{$e$}}_{N}^{H}(x+\triangle x,g)-\mbox{\boldmath{$e$}}_{N}(x,g)=\widehat{\mbox{\boldmath{$\iGamma$}}}
_{N\mu }(x,g)\triangle x^{\mu }+O(\triangle x)^{2}.
\end{equation}

\noindent
In the same way we have the covariant difference of $\mbox{\boldmath{$e$}}_{N}$ along $\mbox{\boldmath{$Z$}}_2$%
\begin{equation}
\triangle _{r}\mbox{\boldmath{$e$}}_{N}(x,g)\equiv \mbox{\boldmath{$e$}}_{N}^{H}(x,g+r)-\mbox{\boldmath{$e$}}_{N}(x,g)=\widehat{\mbox{\boldmath{$\iGamma$}}}%
_{Nr}(x,g)\triangle ^{r}z(g),
\end{equation}

\noindent
where
\begin{equation}
\widehat{\mbox{\boldmath{$\iGamma$}}}_{Nr}(x,g)\equiv \mbox{\boldmath{$e$}}_{M}(x,g)\widehat{\iGamma }_{\
Nr}^{M}(x,g).
\end{equation}

\noindent
In the previous paper\cite{Kokado} we have shown that the rotation matrix $H_{\
N}^{M}(x,g,g+r)$ has the form, $H_{\ N}^{M}(x,g,g+r)$ $=\delta _{\
N}^{M}+\widehat{\iGamma }_{\ Nr}^{M}(x,g)\triangle ^{r}z(g)$, having no
term of $O(\triangle ^{r}z)^{2}$, $i.e.$, its Taylor expansion is cut off
only up to the first order of $\triangle ^{r}z$. This sharply differs from $%
H_{\ N}^{M}(x,x+\triangle x,g)$, which is expanded into an infinite
power series of $\triangle x$. (See Appendix.)

We are now in a position to consider the torsion on $M_{4}\times \mbox{\boldmath{$Z$}}_2$. 
There are three kinds of torsion in this space. They will be shown to be

\begin{eqnarray}
\mbox{\boldmath{$T$}}^{(1)} &=&[\triangle _{1x},\triangle _{2x}]O(x,g)=
[\widehat{\mbox{\boldmath{$\iGamma$}}}_{\nu \mu }(x,g)-
\widehat{\mbox{\boldmath{$\iGamma$}}}_{\mu \nu }(x,g)]\triangle _{1}x^{\mu }\triangle
_{2}x^{\nu }, \\
\mbox{\boldmath{$T$}}^{(2)} &=&[\triangle _{x},\triangle _{r}]O(x,g)=
[\widehat{\mbox{\boldmath{$\iGamma$}}}_{r\mu }(x,g)-
\widehat{\mbox{\boldmath{$\iGamma$}}}_{\mu r}(x,g)]\triangle x^{\mu }\triangle ^{r}z(g),
\\
\mbox{\boldmath{$T$}}^{(3)} &=&[\triangle _{r}\triangle _{r}+2\triangle _{r}]O(x,g)=
-\widehat{\mbox{\boldmath{$\iGamma$}}}_{rr}(x,g)(\triangle ^{r}z(g))^{2}.
\end{eqnarray}

\newlength{\minitwocolumn}
\setlength{\minitwocolumn}{.5\textwidth}
\addtolength{\minitwocolumn}{-1.0\columnsep}
\ \\
\begin{minipage}[t]{\minitwocolumn}
The first torsion $\mbox{\boldmath{$T$}}^{(1)}$ is known to be of the conventional type. For the
latter convenience let us derive this formula. First we consider two
sequential mappings of the origin $O(P_{3})$ from $T(P_{3})$ onto $T(P)$
along two paths $C_{1}$ and $C_{2}$ depicted in Fig.1, where coordinates of $%
P$, $P_{1}$, $P_{2}$ and $P_{3}$ are \\ 
\end{minipage}
\hspace{\columnsep}
\begin{minipage}[t]{\minitwocolumn}
\begin{center}
\ \vspace{0ex}\\
\setlength{\unitlength}{1mm}
\begin{picture}(75,35)(-6,-7)
 \put(30,5){\vector(-1,1){6}}\put(24,11){\line(-1,1){4}}
 \put(20,15){\vector(1,1){6}}\put(26,21){\line(1,1){4}}
 \put(30,5){\vector(1,1){6}}\put(36,11){\line(1,1){4}}
 \put(40,15){\vector(-1,1){6}}\put(34,21){\line(-1,1){4}}
 \put(30,5){\circle*{1}}
 \put(20,15){\circle*{1}}
 \put(40,15){\circle*{1}}
 \put(30,25){\circle*{1}}
 \put(29,27){$P$}
 \put(14,14){$P_1$}
 \put(42,14){$P_2$}
 \put(29,1){$P_3$}
 \put(19,20){$C_1$}
 \put(36,20){$C_2$}
 \put(25,-6){{\bf Fig.1.}}
\end{picture}
\vspace{0ex}\\
\end{center}
\end{minipage}

\begin{eqnarray}
P&=&(x,g), \nonumber \\
P_{1}&=&(x+\triangle _{1}x,g), \nonumber \\
P_{2}&=&(x+\triangle _{2}x,g), \nonumber \\
P_{3}&=&(x+\triangle _{1}x+\triangle _{2}x,g). \nonumber 
\end{eqnarray}

\noindent
They are given by 
\begin{equation}
C_{1}=U(P,P_{1})U(P_{1},P_{3})O(P_{3}),
\end{equation}
\begin{equation}
C_{2}=U(P,P_{2})U(P_{2},P_{3})O(P_{3}).
\end{equation}

\noindent
The difference between $C_{1}$ and $C_{2}$ gives just the torsion 
\begin{equation}
\mbox{\boldmath{$T$}}^{(1)}=C_{1}-C_{2}.
\end{equation}

\noindent
Noting 
\begin{equation}
\triangle _{2x}O(P)=U(P,P_{2})O(P_{2})-O(P)=\mbox{\boldmath{$e$}}_{\nu }(x,g)\triangle
_{2}x^{\nu },
\end{equation}

\noindent
and from Eq. (2.9) one gets 
\begin{eqnarray}
\triangle _{1x}\triangle _{2x}O(P)
&=&U(P,P_{1})U(P_{1},P_{3})O(P_{3})-U(P,P_{1})O(P_{1}) \nonumber \\
& &-U(P,P_{2})O(P_{2})+O(P) \nonumber \\
&=&\triangle _{1x}\mbox{\boldmath{$e$}}_{\nu }(x,g)\triangle _{2}x^{\nu }  \nonumber \\
&=&\widehat{\mbox{\boldmath{$\iGamma$}}}_{\nu \mu }(x,g)\triangle _{1}x^{\mu }\triangle
_{2}x^{\nu }. 
\end{eqnarray}

\noindent
Thus we find 
\begin{equation}
(\triangle _{1x}\triangle _{2x}-\triangle _{2x}\triangle
_{1x})O(P)=[U(P,P_{1})U(P_{1},P_{3})-U(P,P_{2})U(P_{2},P_{3})]O(P_{3})=\mbox{\boldmath{$T$}}^{(1)}.
\end{equation}

\noindent
Namely, $[\triangle _{1x},\triangle _{2x}]O(P)$ gives just the first torsion 
$\mbox{\boldmath{$T$}}^{(1)}$, (2.12).
\ \\
\begin{minipage}[t]{\minitwocolumn}
\ \\
The second torsion $\mbox{\boldmath{$T$}}^{(2)}$ will be derived in the same way as above if we
consider two mappings of the origin $O(x+\triangle x,g+r)$ from $%
T(x+\triangle x,g+r)$ onto $T(x,g)$ along two paths $C_{3}$ and $C_{4}$
depicted in Fig.2. They are given by \\ 
\end{minipage}
\hspace{\columnsep}
\begin{minipage}[t]{\minitwocolumn}
\begin{center}
\ \vspace{-1ex}\\
\setlength{\unitlength}{1mm}
\begin{picture}(75,35)(-10,-7)
 \put(10,5){\line(1,0){10}}
 \put(10,20){\line(1,0){10}}
 \put(35,5){\line(1,0){10}}
 \put(35,20){\line(1,0){10}}
 \put(20,5){\vector(0,1){8}}\put(20,13){\line(0,1){7}}
 \put(35,20){\vector(-1,0){8}}\put(27,20){\line(-1,0){7}}
 \put(35,5){\vector(-1,0){8}}\put(27,5){\line(-1,0){7}}
 \put(35,5){\vector(0,1){8}}\put(35,13){\line(0,1){7}}
 \put(20,5){\circle*{1}}
 \put(20,20){\circle*{1}}
 \put(35,5){\circle*{1}}
 \put(35,20){\circle*{1}}
 \put(19,1){$x$}
 \put(31,1){$x+\iDelta x$}
 \put(19,22){$x$}
 \put(31,22){$x+\iDelta x$}
 \put(50,4){$g+r$}
 \put(50,19){$g$}
 \put(24,22){$C_3$}
 \put(15,11){$C_4$}
 \put(22,-6){{\bf Fig.2.}}
\end{picture} 
\vspace{-1ex}\\
\end{center}
\end{minipage}
\begin{eqnarray}
C_{3}&=&U(x,x+\triangle x,g)U(x+\triangle x,g,g+r)O(x+\triangle x,g+r), \\
C_{4}&=&U(x,g,g+r)U(x,x+\triangle x,g+r)O(x+\triangle x,g+r).
\end{eqnarray}

\noindent
The difference between $C_{3}$ and $C_{4}$ gives the second torsion $%
\mbox{\boldmath{$T$}}^{(2)}. $ After the similar calculation as $\mbox{\boldmath{$T$}}^{(1)}$ we have 
\begin{eqnarray}
\mbox{\boldmath{$T$}}^{(2)} &=&C_{3}-C_{4}  \nonumber \\
&=&(\triangle _{x}\triangle _{r}-\triangle _{r}\triangle _{x})O(x,g),
\end{eqnarray}

\noindent
which is just Eq.(2.13).
\ \\
\begin{minipage}[t]{\minitwocolumn}
\ \\
 Finally the third torsion $\mbox{\boldmath{$T$}}^{(3)}$ will be obtained if we consider the
mappings of the origin $O(x,g)$ from $T(x,g)$ onto $T(x,g+r)$ and again onto
the same $T(x,g)$. The paths of the mappings are depicted in Fig.3. The
difference between the mapped point and $O(x,g)$ gives the torsion \\
\end{minipage}
\hspace{\columnsep}
\begin{minipage}[t]{\minitwocolumn}
\begin{center}
\ \vspace{-1ex}\\
\setlength{\unitlength}{1mm}
\begin{picture}(75,35)(-10,-8)
 \put(25,5){\line(1,0){10}}
 \put(25,20){\line(1,0){10}}
 \put(35,5){\line(1,0){10}}
 \put(35,20){\line(1,0){10}}
 \put(36,5){\vector(0,1){8}}\put(36,13){\line(0,1){7}}
 \put(35,20){\vector(0,-1){8}}\put(35,12){\line(0,-1){7}}
 \put(35.5,5){\circle*{1}}
 \put(35.5,20){\circle*{1}}
 \put(30,1){$(x,g+r)$}
 \put(30,22){$(x,g)$}
 \put(30,-7){{\bf Fig.3.}}
\end{picture}
\vspace{-1ex}\\
\end{center}
\end{minipage}
\begin{equation}
\mbox{\boldmath{$T$}}^{(3)}=U(x,g,g+r)U(x,g+r,g)O(x,g)-O(x,g).
\end{equation}

\noindent
In order to obtain the explicit formula for $\mbox{\boldmath{$T$}}^{(3)}$, let us take the
covariant difference of both sides of (2.2) 
\begin{eqnarray}
\triangle _{r}\triangle _{r}O(x,g) &=&\triangle
_{r}[U(x,g,g+r)O(x,g+r)-O(x,g)] \nonumber \\
&=&U(x,g,g+r)[U(x,g+r,g)O(x,g)-O(x,g+r)]  \nonumber \\
&&-U(x,g,g+r)O(x,g+r)+O(x,g)  \nonumber \\
&=&U(x,g,g+r)U(x,g+r,g)O(x,g)  \nonumber \\
&&-2U(x,g,g+r)O(x,g+r)+O(x,g). 
\end{eqnarray}

\noindent
Hence we have 
\begin{equation}
(\triangle _{r}\triangle _{r}+2\triangle
_{r})O(x,g)=U(x,g,g+r)U(x,g+r,g)O(x,g)-O(x,g)=\mbox{\boldmath{$T$}}^{(3)}.
\end{equation}

\noindent
Namely, the left-hand side quantity is equal to the third torsion. From the
last equation of (2.2) and (2.10) we find 
\begin{eqnarray}
\triangle _{r}\triangle _{r}O(x,g) &=&\triangle _{r}[\mbox{\boldmath{$e$}}_{r}(x,g)\triangle
^{r}z(g)] \nonumber \\
&=&\mbox{\boldmath{$e$}}_{r}^{H}(x,g+r)\triangle ^{r}z(g+r)-\mbox{\boldmath{$e$}}_{r}(x,g)\triangle ^{r}z(g) 
\nonumber \\
&=&[\mbox{\boldmath{$e$}}_{r}(x,g)+\widehat{\mbox{\boldmath{$\iGamma$}}}_{rr}(x,g)\triangle ^{r}z(g)]\triangle
^{r}z(g+r)-\mbox{\boldmath{$e$}}_{r}(x,g)\triangle ^{r}z(g)  \nonumber \\
&=&\widehat{\mbox{\boldmath{$\iGamma$}}}_{rr}(x,g)\triangle ^{r}z(g)\triangle
^{r}z(g+r)+\mbox{\boldmath{$e$}}_{r}(x,g)[\triangle ^{r}z(g+r)-\triangle ^{r}z(g)]  \nonumber \\
&=&-\widehat{\mbox{\boldmath{$\iGamma$}}}_{rr}(x,g)(\triangle ^{r}z(g))^{2}-2\mbox{\boldmath{$e$}}_{r}(x,g)\triangle
^{r}z(g).  
\end{eqnarray}

\noindent
Here we have used $\triangle ^{r}z(g+r)=z(g)-z(g+r)=-\triangle ^{r}z(g)$. Thus, finally we get 
\begin{equation}
\mbox{\boldmath{$T$}}^{(3)}=-\widehat{\mbox{\boldmath{$\iGamma$}}}_{rr}(x,g)(\triangle ^{r}z(g))^{2}.
\end{equation}

The first and second torsions vanish when the affine connection
 $\widehat{\mbox{\boldmath{$\iGamma$}}}
_{MN}(x,g)$ is symmetric with respect to $M$ and $N$, $\widehat{\mbox{\boldmath{$\iGamma$}}} 
_{MN}=\widehat{\mbox{\boldmath{$\iGamma$}}}
_{NM}$. However, the third torsion $\mbox{\boldmath{$T$}}^{(3)}$ remains generally finite.
\newpage

%%%%%%%%%%%%%%%%%%%%%%%%%%%%%% Section 3 %%%%%%%%%%%%%%%%%%%%%%%%%%%%%%%
\setcounter{equation}{0}
\section{The isometry condition}
\indent

 The manifold $M_{4}\times \mbox{\boldmath{$Z$}}_2$ can be regarded as the Kaluza-Klein like
space where the fifth continuous dimension is replaced by two discrete
points $z(e)$ and $z(r)$. The line element $\triangle s$ of this space is assumed
to be 
\begin{eqnarray}
\triangle s^{2} &=&g_{\mu \nu }\triangle x^{\mu }\triangle x^{\nu }+\lambda
^{2}(x)\triangle z^{2} \nonumber \\
&=&G_{MN}(x)\triangle x^{M}\triangle x^{N},  
\end{eqnarray}

\noindent
where 
\begin{equation}
\triangle x^{N}=(\triangle x^{\mu },\triangle x^{r}\equiv \triangle
z=z(r)-z(e)),
\end{equation}

\noindent
and $G_{MN}(x)$ is regarded as the five-dimensional metric of $M_{4}\times
\mbox{\boldmath{$Z$}}_2$.

\noindent
Here we have considered a simple case that the four-dimensional metric $%
g_{\mu \nu }(x)$ and the scalar field $\lambda (x)$ are independent of $z(g)$
on $\mbox{\boldmath{$Z$}}_2$ and are functions only of $x\in M_{4}$. The first and second
kinds of torsions $\mbox{\boldmath{$T$}}^{(1)}$ and $\mbox{\boldmath{$T$}}^{(2)}$ are assumed to be zero, $i.e.$, 
\begin{equation}
\widehat{\iGamma }_{\ \mu \nu }^{N}(x,g)=\widehat{\iGamma }_{\ \nu \mu
}^{N}(x,g),\mbox{ \ }\widehat{\iGamma }_{\ \mu r}^{N}(x,g)=\widehat{%
\iGamma }_{\ r\mu }^{N}(x,g),
\end{equation}

\noindent
whereas the third kind of torsion $\mbox{\boldmath{$T$}}^{(3)}$ is not necessarily vanished 
\begin{equation}
\widehat{\iGamma }_{\ rr}^{N}(x,g)\neq 0.
\end{equation}

The metric is defined by the inner product 
\begin{equation}
G_{MN}(x,g)=G_{MN}(x)=\mbox{\boldmath{$e$}}_{M}(x,g)\cdot \mbox{\boldmath{$e$}}_{N}(x,g).
\end{equation}

\noindent
Let the manifold $M_{4}\times \mbox{\boldmath{$Z$}}_2$ be isometric, that is, any inner
product of vectors is invariant under the parallel-transportation or the
mapping of vectors. Then we have 
\begin{equation}
G_{MN}(x+\triangle x,g)=\mbox{\boldmath{$e$}}_{M}(x+\triangle x,g)\cdot \mbox{\boldmath{$e$}}_{N}(x+
\triangle x,g)=\mbox{\boldmath{$e$}}_{M}^{H}(x+\triangle x,g)\cdot \mbox{\boldmath{$e$}}_{N}^{H}(x+\triangle x,g).
\end{equation}

\noindent
Substituting (2.9) into (3.6) this is reduced to 
\begin{equation}
\partial _{\lambda }G_{MN}=\widehat{\iGamma }_{M\lambda N}+
\widehat{\iGamma }_{N\lambda M},\mbox{ \ } \widehat{\iGamma }_{M\lambda N}\equiv G_{MK}\widehat{
\iGamma }^{K}_{\ \lambda N}.
\end{equation}

\noindent
In the same way by using (2.10) for the direction to $\mbox{\boldmath{$Z$}}_2$ we get
\begin{equation}
\partial _{r}G_{MN}=\widehat{\iGamma }_{MrN}+\widehat{\iGamma }%
_{NrM}+\widehat{\iGamma }_{KrM}\widehat{\iGamma }^{K}_{rN}\triangle ^{r}z.
\end{equation}

\noindent
Since any function $f(z(g))$ on $\mbox{\boldmath{$Z$}}_2$ is a linear function of $z(g)$
(see (1.5)), the derivative on $\mbox{\boldmath{$Z$}}_2$, $\partial _{r}f\equiv \partial f/\partial
z(g)$, can be defined without taking the limit $\triangle
^{r}z(g)\rightarrow 0$ and is always independent of $g$. Namely, Eq.(3.8) is
valid for any $\triangle ^{r}z$. Of course, the left-hand side of (3.8) is
zero from the assumption that $G_{MN}$ is independent of $g$.

Noting $\partial _{r}\triangle ^{r}z(g)=-2$ and differentiating the
right-hand side of (3.8) with respect to $z(g)$, we have 
\begin{equation}
0=\partial _{r}\widehat{\iGamma }_{MrN}+\partial _{r}\widehat{\iGamma }_{NrM}-2%
\widehat{\iGamma }_{KrM}\widehat{\iGamma }^{K}_{\ rN}+\partial _{r}(%
\widehat{\iGamma }_{KrM}\widehat{\iGamma }^{K}_{\ rN})\triangle ^{r}z.
\end{equation}

\noindent
Taking the limit $\triangle ^{r}z\rightarrow 0$ we get 
\begin{equation}
(\partial _{r}\widehat{\iGamma }_{MrN}+\partial _{r}\widehat{\iGamma }%
_{NrM})_{0}=2\iGamma _{KrM}\iGamma _{\ rN}^{K},\mbox{\ (%
}\triangle
^{r}z=0)
\end{equation}

\noindent
where the suffix $_{0}$ and $\iGamma $ without the hat show that these
quantities are independent of $g$. In the same limit $\triangle
^{r}z\rightarrow 0$ Eqs.(3.7) and (3.8) tend to 
\begin{equation}
\partial _{\lambda }G_{MN}=\iGamma _{M\lambda N}+\iGamma _{N\lambda M},\mbox{%
\ (}\triangle ^{r}z=0)
\end{equation}
\begin{equation}
\partial _{r}G_{MN}=\iGamma _{MrN}+\iGamma _{NrM},\mbox{\ (}\triangle
^{r}z=0)
\end{equation}

\noindent
Note that from (3.8) or (3.9) one cannot derive $\widehat{\iGamma }_{KrM}%
\widehat{\iGamma }^{K}_{rN}=0$ or $\partial _{r}(\widehat{\iGamma }_{KrM}%
\widehat{\iGamma }^{K}_{rN})=0$ because the other terms contain $\triangle
^{r}z$.\footnote[1]{Since any function $f(g)$ on $\mbox{\boldmath{$Z$}}_2$ has 
a form, $f(g)=A+Bz(g)$, we have
\begin{eqnarray*}
f(e) &=&\frac{1}{2}[f(e)+f(r)]+\frac{1}{2}[f(e)-f(r)]=\overline{f}-\frac{1}{2%
}B\triangle z(e),\triangle z(e)=z(r)-z(e), \\
f(r) &=&\overline{f}-\frac{1}{2}B\triangle z(r),\triangle z(r)=z(e)-z(r),
\end{eqnarray*}

so that
\[
f(g)=\overline{f}-\frac{1}{2}B\triangle ^{r}z(g),\triangle
^{r}z(g)=z(g+r)-z(g).
\]

Namely, the function $f(g)$ contains $\triangle ^{r}z(g)$.
}
 From (3.11) and (3.12) and from (3.3) one obtains the expression of $%
\iGamma _{LMN}$ in terms of $G_{MN}$%
\begin{equation}
\iGamma _{LMN}=\frac{1}{2}(\partial _{M}G_{LN}+\partial _{N}G_{LM}-\partial
_{L}G_{MN}).
\end{equation}

\noindent
Since the metric $G_{MN}$ is given by (3.1) we have 
\begin{eqnarray}
\iGamma _{\lambda \mu \nu }&=&\frac{1}{2}(\partial _{\mu }g_{\lambda \nu
}+\partial _{\nu }g_{\lambda \mu }-\partial _{\lambda }g_{\mu \nu }),  \\
\iGamma _{rr\mu }&=&-\iGamma _{\mu rr}=\lambda \partial _{\mu }\lambda , \\
\iGamma _{r\mu \nu }&=&\iGamma _{\mu \nu r}=\iGamma _{\mu r\nu }=0, \\
\iGamma _{rrr}&=&0.
\end{eqnarray}

On the other hand, from (3.10) with $M=\mu $ and $N=\nu $ and from (3.16) we
see 
\begin{equation}
(\partial _{r}\widehat{\iGamma }_{\mu r \nu }+\partial _{r}\widehat{\iGamma }%
_{\nu r\mu })_{0}=2\iGamma _{Kr\mu }\iGamma _{\ r\nu }^{K}=2\iGamma _{%
rr\mu }\iGamma _{\ r\nu }^{r}%
,
\end{equation}

\noindent
and from (3.7) 
\begin{equation}
\partial _{\nu }G_{\mu r}=\widehat{\iGamma }_{\mu \nu r}+\widehat{\iGamma }_{r\nu \mu }=0.
\end{equation}

\noindent
The equation (3.19) yields 
\[
\widehat{\iGamma }_{\mu \nu r}=-\widehat{\iGamma }_{r\nu \mu } \nonumber 
\]

\noindent
and 
\[
\widehat{\iGamma }_{\nu \mu r}=-\widehat{\iGamma }_{r\mu \nu }=-\widehat{%
\iGamma }_{r\nu \mu },
\]

\noindent
so that 
\begin{equation}
\widehat{\iGamma }_{\mu \nu r}=\widehat{\iGamma }_{\nu \mu r}.
\end{equation}

\noindent
From (3.20) one can see that the first and second terms in (3.18) are equal
to each other, hence we have 
\begin{equation}
(\partial _{r}\widehat{\iGamma }_{\mu \nu r})_{0}=\iGamma _{rr\mu }\iGamma _{\
r\nu }^{r}=-(\partial _{r}\widehat{\iGamma }_{r\mu \nu })_{0}.
\end{equation}

Other useful relations come from (3.10) 
\begin{equation}
(\partial _{r}\widehat{\iGamma }_{rrr})_{0}=\iGamma _{Krr}\iGamma _{\
rr}^{K}=\iGamma _{\rho rr}\iGamma _{\ rr}^{\rho },
\end{equation}
\begin{equation}
(\partial _{r}\widehat{\iGamma }_{\mu rr}+\partial _{r}\widehat{\iGamma }%
_{rr\mu })_{0}=2\iGamma _{Kr\mu }\iGamma _{\ rr}^{K}=0.
\end{equation}

\noindent
In the next section we use Eqs.(3.14)-(3.17) and (3.21) -(3.23) to calculate
the curvature.
\newpage

%%%%%%%%%%%%%%%%%%%%%%%%%%%%%% Section 4 %%%%%%%%%%%%%%%%%%%%%%%%%%%%%%%
\setcounter{equation}{0}
\section{Curvature}

\indent

Corresponding to three kinds of torsions there are also three kinds of
curvature tensors. They are given by 
\begin{eqnarray}
\lbrack \triangle _{1x},\triangle _{2x}]\mbox{\boldmath{$e$}}_{N}&=&\mbox{%
\boldmath{$e$}}_{K}R_{\ N\mu \nu }^{K}\Delta _{1}x^{\mu }\Delta _{2}x^{\nu },
 \nonumber \\
R_{\ N\mu \nu }^{K}&=&\partial _{\mu }\widehat{\iGamma }_{\ N\nu
}^{K}-\partial _{\nu }\widehat{\iGamma }_{\ N\mu }^{K}+\widehat{\iGamma }%
_{\ J\mu }^{K}\widehat{\iGamma }_{\ N\nu }^{J}-\widehat{\iGamma }%
_{\ J\nu }^{K}\widehat{\iGamma }_{\ N\mu }^{J}, \\
\lbrack \triangle _{x},\triangle _{r}]\mbox{\boldmath{$e$}}_{N}&=&\mbox{%
\boldmath{$e$}}_{K}R_{\ N\mu r}^{K}\Delta x^{\mu }\Delta ^{r}z, \nonumber \\ 
R_{\ N\mu r}^{K}&=&\partial _{\mu }\widehat{\iGamma }_{\
Nr}^{K}-\partial _{r}\widehat{\iGamma }_{\ N\mu }^{K}+\widehat{\iGamma }%
_{\ J\mu }^{K}\widehat{\iGamma }_{\ Nr}^{J}-\widehat{\iGamma }_{\
Jr}^{K}\widehat{\iGamma }_{\ N\mu }^{J}, \\
(\triangle _{r}\triangle _{r}+2\triangle _{r})\mbox{\boldmath{$e$}}_{N}&=&\mbox{%
\boldmath{$e$}}_{K}R_{\ Nrr}^{K}(\Delta ^{r}z)^{2}, \nonumber \\
R_{\ Nrr}^{K}&=&-\partial _{r}\widehat{\iGamma }_{\ Nr}^{K}-\widehat{%
\iGamma }_{\ Jr}^{K}\widehat{\iGamma }_{\ Nr}^{J}.
\end{eqnarray}

\noindent
Their geometrical meanings have been previously clarified \cite{Kokado}. In the limit $%
\Delta ^{r}z\rightarrow 0$, by using Eqs.(3.11) and (3.12) three curvature
tensors above can be rewritten in forms 
\begin{eqnarray}
R_{MN\mu \nu }&=&\partial _{\mu }\iGamma _{MN\nu }-\partial
_{\nu }\iGamma _{MN\mu }-\iGamma _{J\mu M}\iGamma _{\
N\nu }^{J}+\iGamma _{J\nu M}\iGamma _{\ N\mu }^{J}, \\
R_{MN\mu r}&=&\partial _{\mu }\iGamma _{MNr}-(\partial
_{r}\widehat{\iGamma }_{MN\mu }) _{0}-\iGamma _{J\mu M}\iGamma _{\
Nr}^{J}+\iGamma _{JrM}\iGamma _{\ N\mu }^{J}, \\
R_{MNrr}&=&-(\partial _{r}\widehat{\iGamma }_{MNr})_{0}+\iGamma_{\
JrM}\iGamma _{\ Nr}^{J}.
\end{eqnarray}

We first note from (4.6) and (3.22)
\begin{eqnarray}
R_{rrrr} &=&-(\partial _{r}\widehat{\iGamma }_{rrr})_{0}+
\iGamma _{Jrr}\iGamma _{\ rr}^{J} \nonumber \\
&=&-(\partial _{r}\widehat{\iGamma }_{rrr})_{0}+\iGamma _{\rho rr}\Gamma _{%
\ rr}^{\rho } \nonumber \\
&=&0, \nonumber 
\end{eqnarray}

\noindent
hence 
\begin{equation}
R_{\  rrr}^{r}=G^{rr}R_{rrrr}=0.
\end{equation}

\noindent
From (4.5) and (3.21) the relevant component $R_{\nu r\mu r}$ is reduced to 
\begin{eqnarray}
R_{\nu r\mu r} &=&\partial _{\mu }\iGamma _{\nu rr}-(\partial _{r}\widehat{%
\iGamma }_{\nu r\mu })_{0}-\iGamma _{J\mu \nu }\iGamma _{\ %
rr}^{J}+\iGamma _{Jr\nu }\iGamma _{\ r\mu }^{J}   \nonumber \\
&=&\partial _{\mu }\iGamma _{\nu rr}-\iGamma _{\rho \mu \nu }\iGamma _{\ %
rr}^{\rho }-\iGamma _{rr\nu }\iGamma _{\ r\mu }^{r}+
\iGamma _{rr\nu }\iGamma _{\ r\mu }^{r}  \nonumber \\
&=&\nabla _{\mu }\iGamma _{\nu rr}  \nonumber \\
&=&-\nabla _{\mu }(\lambda \partial _{\nu }\lambda ),
\end{eqnarray}

\noindent
where we have used (3.15) and $\nabla _{\mu }$ is the covariant derivative
in $M_{4}$, hence 
\begin{equation}
R_{\ r\rho r}^{\rho }=-\nabla ^{\rho }(\lambda \partial _{\rho
}\lambda ).
\end{equation}

\noindent
In the same way we get 
\begin{eqnarray}
R_{r\nu r\mu } &=&(\partial _{r}\widehat{\iGamma }_{r\nu \mu })_{0}-
\partial _{\mu }\iGamma _{rr\nu }+\iGamma _{J\mu r}\iGamma _{\ \nu
r}^{J}-\iGamma _{Jrr}\iGamma _{\ \nu \mu }^{J} \nonumber \\
&=&-\iGamma _{rr\nu }\iGamma _{\ r\mu }^{r}-
\partial _{\mu }\iGamma _{rr\nu }+\iGamma _{rr\mu }\iGamma _{\ r\nu }^{r}+
\iGamma _{rr\rho}\iGamma _{\ \mu \nu }^{\rho }  \nonumber \\
&=&-\nabla _{\mu }\iGamma _{rr\nu }  \nonumber \\
&=&-\nabla _{\mu }(\lambda \partial _{\nu }\lambda ), 
\end{eqnarray}

\noindent
so that 
\begin{equation}
R^{r}_{\ \mu r\nu }=G^{rr}R_{r\mu r\nu }=-\frac{1}{\lambda ^{2}}\nabla _{\nu }(\lambda
\partial _{\mu }\lambda ).
\end{equation}

From (4.4) we have
\begin{eqnarray}
R_{\rho \sigma \mu \nu }&=&\partial _{\mu }\iGamma _{\rho \sigma \nu
}-\partial _{\nu }\iGamma _{\rho \sigma \mu }-\iGamma %
_{J\mu \rho }\iGamma _{\ \sigma \nu }^{J}+\iGamma %
_{J\nu \rho }\iGamma _{\ \sigma \mu }^{J} \nonumber \\
&=&\partial _{\mu }\iGamma _{\rho \sigma \nu
}-\partial _{\nu }\iGamma _{\rho \sigma \mu }-\iGamma %
_{\lambda \mu \rho }\iGamma _{\ \sigma \nu }^{\lambda }+\iGamma %
_{\lambda \nu \rho }\iGamma _{\ \sigma \mu }^{\lambda }.
\end{eqnarray}

\noindent
This gives the 4-dimensional conventional Riemann scalar curvature
\begin{equation}
R^{(4)}=g^{\mu \nu }R_{\ \mu \rho \nu }^{\rho },
\end{equation}

\noindent
which is refered to the first kind of scalar curvature. In addition to 
$R^{(4)}$ we have the second and third kinds of scalar curvatures
corresponding to (4.5) and (4.6), respectively. For the second kinds of 
scalar curvatures we have two types, which are defined by
\begin{eqnarray}
R_{2nd}^{(1)}&=&g^{\mu \nu }R_{\ \mu r\nu }^{r}
=-\frac{1}{\lambda ^{2}}\nabla ^{\rho }(\lambda \partial _{\rho
}\lambda ), \\
R_{2nd}^{(2)}&=&G^{rr}R_{\ r\rho r}^{\rho }
=-\frac{1}{\lambda ^{2}}\nabla ^{\rho }(\lambda \partial _{\rho
}\lambda ).
\end{eqnarray}

\noindent
The third kind of scalar curvature is defined by
\begin{equation}
R_{3rd}=G^{rr}R_{\ rrr}^{r}=0,
\end{equation}

\noindent
which vanishes owing to (4.7).

Now, in order to construct the gravity action, we consider that the 
three scalar curvatures $R^{(4)}$, $R_{2nd}^{(1)}$, and $R_{2nd}^{(2)}$
are all scalar quantities on $M_{4}$ and they are never mixed with 
each other under general coordinate transformations. The most simple 
gravity action on $M_{4}\times {\mbox{\boldmath{$Z$}}_2}$ linear to 
$R^{\prime }s$, therefore, should be composed of three terms
\begin{eqnarray}
I&=&\int_{M_{4}}\int_{\mbox{\boldmath{$Z$}}_2}\sqrt{-g}\frac{1}{\lambda ^{2}}
[R^{(4)}+c_{1}R_{2nd}^{(1)}+c_{2}R_{2nd}^{(2)}]  \nonumber \\
&=&\int_{M_{4}}\int_{\mbox{\boldmath{$Z$}}_2}\sqrt{-g}[\frac{1}{\lambda ^{2}}
R^{(4)}-(c_{1}+c_{2})\frac{1}{\lambda ^{4}}\nabla
^{\rho }(\lambda \partial _{\rho }\lambda )].
\end{eqnarray}
where $g\equiv det(g_{\mu \nu })$ and $c_{1}$, $c_{2}$ are real dimensionless
arbitrary constants. From (3.1) the dimension of $\lambda $ is $[\lambda ]
=L(length)$, if $\triangle z$ is dimensionless. Since the action should be dimensionless, 
the scalar curvatures should be multiplied by $1/\lambda ^{2}$ like (4.17) 
because $[R's]=L^{-2}$. After partial-integration for the second term in 
(4.17) and setting $2/\lambda ^{2}=\phi $, the action summed over 
$\mbox{\boldmath{$Z$}}_2$ is reduced to%

\begin{equation}
I =\int_{M_{4}}\sqrt{-g}[\phi R^{(4)}-\omega \frac{\partial ^{\rho }\phi 
\partial _{\rho }\phi }{\phi }],
\end{equation}

\noindent%
where $\omega =c_{1}+c_{2}$. This is nothing but the BD theory with the
arbitrary BD coupling constant $\omega $.
\newpage

%%%%%%%%%%%%%%%%%%%%%%%%%%%%%% Section 5 %%%%%%%%%%%%%%%%%%%%%%%%%%%%%%%
\setcounter{equation}{0}
\section{Concluding remarks}

\indent

On the basis of the equivalence assumption stated in the introduction
and also of the new isometry condition (3.8) we have derived 
the Brans-Dicke theory on the manifold $M_{4}\times \mbox{\boldmath{$Z$}}_2$.
In the previous work\cite{Kokado} we have not taken into account of 
the contribution from the last term in (3.8). 
This term is crucially so important as to eliminate 
$R_{rrrr}$, the Riemann tensor of the third type (4.3), and to yield the BD
kinetic term, which comes out of the Riemann curvature of the second type (4.2).

The BD coupling constant $\omega $ has become to be arbitrary, contrary
to the previous work. This comes from the fact that the three scalar curvatures
$R^{(4)}$, $R_{2nd}^{(1)}$, and $R_{2nd}^{(2)}$ are all independent scalar
quantities on $M_{4}$ and they are never mixed with each other under general 
coordinate transformations. This fact allows us to introduce arbitrary parameters
$c_{1}$, $c_{2}$ into the action, where $\omega $ is given by $\omega = c_{1}+c_{2}$.

We also have clarified the geometrical meaning of torsion in this space.
There are three kinds of torsions, $\widehat{\iGamma }_{\  \mu \nu
}^{M}-\widehat{\iGamma }_{\  \nu \mu }^{M}$, $\widehat{\iGamma }_{%
\  \mu r}^{M}-\widehat{\iGamma }_{\  r\mu }^{M}$ and $-\widehat{%
\iGamma }_{\  rr}^{M}$. In the BD theory first two
torsions should vanish, but the last one remains finite in the limit $%
\triangle ^{r}z\rightarrow 0$, $i.e.$, $-\iGamma _{\  %
rr}^{\mu }=\lambda \partial ^{\mu }\lambda $.

%%%%%%%%%%%%%%%%%%%%%%%%%%%%% acknowledgements %%%%%%%%%%%%%%%%%%%%%%%%%%%%%%%
\setcounter{equation}{0}
\section*{Acknowledgements}

\indent

One of the authors(T.S.) would like to express his thanks to Y. Fujii for his 
useful comments and discussions. Thanks are also due to Z. Maki for his interest 
in this work and for encouraging us.

%%%%%%%%%%%%%%%%%%%%%%%%%%%%%% Appendix A %%%%%%%%%%%%%%%%%%%%%%%%%%%%%%%
\newpage
\renewcommand{\theequation}{\Alph{section}.\arabic{equation}}
\setcounter{equation}{0}
\setcounter{section}{1}
\section*{Appendix}

\indent

More elegant proof of this is given as follows: Let the coordinate $z(g)$
take values $z(e)=a$ and $z(r)=b$. Then we find the identity
\begin{equation}
z^{2}(g)=(a+b)z(g)-ab.
\label{eA01}
\end{equation}

\noindent
From this any polynomial of $z(g)$ is reduced to a linear function of $z(g)$%
. Therefore, any function regular on $z(g)$ can be expressed as (1.5), $i.e.$%
, $f(g)=A+Bz(g)$.

The mapping function $H(g,g+r)=H(z,z^{\prime })$, $z=z(g)$, $z^{\prime
}=z(g+r)$, is also linear to $z$ and $z^{\prime }$, $i.e.$,
\begin{equation}
H(z,z^{\prime })=A+Bz+B^{\prime }z^{\prime }+Czz^{\prime }.
\label{eA02}
\end{equation}

\noindent
Substituting $z^{\prime }=z+\triangle ^{r}z$ into (A.2) and using $%
H(z,z)=1$, we have 
\begin{eqnarray}
H(z,z^{\prime }) &=&1+(B^{\prime }+Cz)\triangle ^{r}z=1+\partial _{z^{\prime
}}H(z,z^{\prime })\mid _{z^{\prime }=z}\triangle ^{r}z  \nonumber \\
&\equiv &1+\widehat{\iGamma }(z)\triangle ^{r}z(g).
\end{eqnarray}

\noindent
Namely, the Taylor expansion of $H(z,z^{\prime })$ is cut off up to the
first order of $\triangle ^{r}z(g)$.

%%%%%%%%%%%%%%%%%%%%%%%%%%%%% references %%%%%%%%%%%%%%%%%%%%%%%%%%%%%%%

\end{document}